\documentclass[aps,amsmath,preprint,tightenlines]{revtex4}
\usepackage{hyperref,graphicx}
\renewcommand{\o}{\omega}
\newcommand{\LNL}{L_{NL}}
\newcommand{\eeqref}[1]{Eq.~(\ref{#1})}

\begin{document}

\title{Decomposing a pulsed optical parametric amplifier into independent squeezers}

\author{A. I.~Lvovsky}
\affiliation{Department of Physics and Astronomy, University of Calgary, Calgary, Alberta T2N 1N4, Canada}
\author{Wojciech Wasilewski}
\affiliation{Institute of Physics, Nicolaus Copernicus University, Grudziadzka 5, 87-100 Toru{\'n}, Poland}
\author{Konrad Banaszek}
\affiliation{Institute of Physics, Nicolaus Copernicus University, Grudziadzka 5, 87-100 Toru{\'n}, Poland}

\date{\today}

\begin{abstract}
We discuss the concept of characteristic squeezing modes applied to a travelling-wave optical parametric amplifier
pumped by an ultrashort pulse. The characteristic modes undergo decoupled single-mode squeezing transformations, and
therefore they form a useful basis to describe the evolution of the entire multimode system. This provides an elegant
and intuitive picture of quantum statistical properties of parametric fluorescence. We analyse the efficiency of
detecting quadrature squeezing, and present results of numerical calculations for a realistic nonlinear medium.
\end{abstract}

\maketitle

\section{Introduction}

Quadrature squeezing \cite{JMOsqueezing} is one of the most accessible optical processes that take us beyond the
semiclassical class of states of electromagnetic radiation, comprising coherent states and their statistical mixtures.
The most elementary theoretical description of squeezing involves just a single radiation mode undergoing evolution
governed by the unitary operator:
\begin{equation}
\label{Eq:SingleModeSqueezing} \hat{U} = \exp\left(\frac{\zeta}{2}[(\hat{a}^\dagger)^2 - \hat{a}^2]\right).
\end{equation}
where $\hat{a}$ is the annihilation operator of the mode, and $\zeta$ is the interaction strength. Such evolution
occurs in the process of parametric down-conversion, which can be realized by means of $\chi^{(2)}$ nonlinearity in
optical media. In this case, the constant $\zeta$ involves the amplitude of the pump beam, the strength of the
nonlinear coupling, and the length of the medium. A common experimental configuration is an optical parametric
amplifier, where a short pump pulse passes once through the medium. This induces squeezing in a mode that is either
initially left empty, or seeded by a coherent beam at half the pump frequency. Obviously,
Eq.~(\ref{Eq:SingleModeSqueezing}) is a very crude approximation to the actual evolution of the system. The realistic
down-conversion is in fact a subtle interplay of the spatio-temporal shape of the pump pulse with the phase-matching
characteristics of the medium for different sets of frequencies and wave vectors. Nevertheless, the basic features of
down-converted light predicted by the single-mode model explain well experimental observations.

The purpose of this paper is to go beyond the single-mode approximation and to discuss theoretically the squeezing
process in a more realistic model, which takes into account the above complexities. This results in multimode
calculations, involving continuous degrees of freedom such as frequency and wave vectors, which can be carried out only
by numerical means. Nevertheless, elementary single-mode squeezing does emerge even in the multimode model. This is a
general consequence of the linearity of the evolution in the field operators and the
associated existence of the Bloch-Messiah reduction \cite{PQ,Bennink,Braunstein}. However, the
modes whose evolution can be described in such a simple way are of a very special form. We shall call them {\em
characteristic squeezing modes} and analyse their properties. The shape of these modes, as well as the strength of the
produced squeezing depend in a nontrivial way on the parameters of the system, including the amplitude and the shape
of the pump pulse as well as the characteristics of the nonlinear medium. The description of realistic squeezing in
terms of characteristic modes provides a simple intuitive picture of the process as well as means to optimize the
amount of detected squeezing.

This paper is organized as follows. First, in Sec.~II we give the theoretical construction of the characteristic
squeezing modes, and motivate it by the well-known Schmidt decomposition of a two-photon wave function. In Sec.~III we
perform the decomposition for a model Gaussian evolution of an optical parametric amplifier, and discuss its
consequences for detecting quadrature squeezing by means of homodyne detection. Sec.~IV presents numerical results for
a down-conversion process realized in a nonlinear waveguide, and points out a simple scaling law for the squeezing
parameters of the characteristic modes. Finally. Sec.~V concludes the paper.

\section{Decomposition}
We will assume here a one-dimensional evolution of the optical field, corresponding to propagation in a single-mode
nonlinear waveguide. Let us first develop an intuition using the perturbative regime, extensively discussed in previous
works in the context of spectral properties of biphotons \cite{KellerRubin,GriceWalmsley}. If the field at the
down-converted frequencies is initially in the vacuum state $|\text{vac}\rangle$, the output state in the limit of weak
pumping can be written as:
\begin{equation}
\hat{U}|\text{vac}\rangle \approx |\text{vac}\rangle + \frac{1}{2} \int \text{d}\omega \int\text{d}\omega' \,
\Psi(\omega, \omega') \hat{a}^\dagger(\omega) \hat{a}^\dagger(\omega') |\text{vac}\rangle.
\end{equation}
The two-photon wave function $\Psi(\omega, \omega')$ is given by a product of of the spectral pump amplitude at
$\omega+\omega'$ and the phase matching function at the down-converted frequencies. For a degenerate process, the wave
function is symmetric with respect to the permutation of its arguments. Consequently, its Schmidt decomposition
\cite{LawWalmsleyEberly} can be written in the form:
\begin{equation}
\Psi(\omega, \omega') = \sum_{n=0}^{\infty} \zeta_n \psi_n^\ast(\omega) \psi_n^\ast(\omega')
\end{equation}
where $\zeta_n$ are non-negative Schmidt coefficients, while $\psi_n^\ast(\omega)$ are mutually orthogonal Schmidt
functions, normalized to one. We have used here complex-conjugated Schmidt functions $\psi_n^\ast(\omega)$ in order to
achieve correspondence with the general case discussed below. The Schmidt decomposition performed above suggests an
introduction of another set of field operators defined by:
\begin{equation}
\label{Eq:SchmidtModes}
\hat{b}_n = \int \text{d}\omega \, \psi_n(\omega) \hat{a}(\omega)
\end{equation}
These operators satisfy standard bosonic commutation relations:
\begin{equation}
[\hat{b}_m, \hat{b}_n]=0, \qquad [\hat{b}_m, \hat{b}_n^\dagger] = \delta_{mn}
\end{equation}
which follows directly from the fact that the Schmidt functions $\psi_n(\omega)$ are orthonormal. The final wave function can now be written as:
\begin{equation}
\label{Eq:Ubmodes} \hat{U}|\text{vac}\rangle \approx |\text{vac}\rangle + \frac{1}{2} \sum_{n=0}^{\infty} \zeta_n (
\hat{b}_n^\dagger )^2 |\text{vac}\rangle.
\end{equation}
It is seen that this wave function could be obtained from the perturbative expansion of the multimode squeezing operator of the form:
\begin{equation}
\hat{U} = \bigotimes_{n=0}^{\infty}
\exp\left(\frac{\zeta_n}{2}[(\hat{b}_n^\dagger)^2 - \hat{b}_n^2]\right).
\end{equation}
As the modes $\hat{b}_n$ are mutually orthogonal, the evolution is effectively decoupled in the modes defined by the
Schmidt functions, and if at the end we detect only one of these modes, then the evolution can be effectively treated
in the single-mode approach.

It turns out that the above decomposition exists not only in the perturbative regime, but, in a generalized form, also
for arbitrarily intense multimode evolution as long as it remains linear in the field operators
\cite{PQ,Bennink,Braunstein}. The non-perturbative case is most conveniently handled in the Heisenberg picture, in
which the final field operators are expressed in terms of the input ones. If the linearity condition is satisfied, the
output field operators $\hat{a}^{\text{out}}(\omega)$ can be in general written in terms of the input field operators
$\hat{a}^{\text{in}}(\omega)$ as a Bogoliubov transformation:
\begin{equation}
\label{Eq:Bogoliubov} \hat{a}^{\text{out}}(\omega) = \int \text{d}\omega' \, \left( C(\omega,\omega')
\hat{a}^{\text{in}}(\omega') + S(\omega,\omega') [\hat{a}^{\text{in}}(\omega')]^\dagger \right)
\end{equation}
where the integral kernels $C(\omega,\omega')$ and $S(\omega,\omega')$ are the Green functions of the equations of
motion for a concrete arrangement. In the perturbative regime, $C(\omega,\omega')= \delta(\omega-\omega')$ and the
function $S(\omega,\omega')$ is equal (up to a certain phase factor) to the biphoton spectrum amplitude
$\Psi(\omega,\omega')$. Of course, the output operators must satisfy the same relations as the input field operators.
This imposes the following conditions on the kernels:
\begin{eqnarray}
\int \text{d}\omega'' [C(\omega,\omega'') S(\omega',\omega'')
- C(\omega',\omega'') S(\omega,\omega'')] & = & 0, \nonumber  \\
\int \text{d}\omega'' [C(\omega,\omega'') C^\ast(\omega',\omega'') - S(\omega,\omega'') S^\ast(\omega',\omega'')] & = &
\delta(\omega-\omega').
\end{eqnarray}
Because of these constraints, the singular-value expansions of the Green functions are not independent, but can be
written using a joint set of functions and parameters as:
\begin{eqnarray}
C(\omega,\omega') & = & \sum_{n=0}^{\infty} \cosh \zeta_n
\psi_n^\ast (\omega ) \phi_n(\omega') \nonumber \\
S(\omega,\omega') & = & \sum_{n=0}^{\infty} \sinh \zeta_n
\psi_n^\ast (\omega ) \phi_n^\ast(\omega').
\end{eqnarray}
Here $\phi_n(\omega)$ and $\psi_n(\omega)$ are two orthonormal sets of functions, and $\zeta_n$ are real parameters.
This decomposition is known as the Bloch-Messiah reduction \cite{Braunstein}. Analogously to
Eq.~(\ref{Eq:SchmidtModes}), we can now define discrete sets of field operators according to:
\begin{eqnarray}
\hat{b}_n^{\text{in}} & = & \int \text{d}\omega \, \phi_n(\omega)
\hat{a}^{\text{in}}(\omega) \nonumber \\
\hat{b}_n^{\text{out}} & = & \int \text{d}\omega \, \psi_n(\omega)
\hat{a}^{\text{out}}(\omega).
\end{eqnarray}
The Bogoliubov transformation introduced in Eq.~(\ref{Eq:Bogoliubov}) induces
now the following transformation of the discrete field operators:
\begin{equation}
\label{Eq:bDecoupled} \hat{b}_n^{\text{out}} = \hat{b}_n^{\text{in}} \cosh \zeta_n +
(
\hat{b}_n^{\text{in}} )^\dagger \sinh \zeta_n .
\end{equation}
It is seen that the evolution is completely decoupled between modes characterized by different indices $n$.
Eq.~(\ref{Eq:bDecoupled}) is analogous to that of single-mode squeezing, and it generalizes the perturbative result
derived in Eq.~(\ref{Eq:Ubmodes}) to the regime of multiple pair generation. Let us note that in general the shape of
the input modes can be different from the shape of the corresponding output mode. As shown in Ref.~\cite{WasilewskiXXX05}, with a
non-chirped pump pulse, the input and output squeezing eigenmodes obey a relation $\psi_n(\omega)=\phi_n^*(\omega)$,
i.e.\ are the exact time reverse of each other.

\section{Homodyne detection}

Let us first discuss consequences of the multimode character of squeezing for homodyne detection with the local
oscillator field described by a certain mode function $\psi_{LO}(\omega)$. In order to expose the essence of the argument, we
will assume the integration kernels in the following Gaussian form:
\begin{eqnarray}
C(\omega,\omega') & = & \delta(\omega - \omega'), \nonumber \\
S(\omega,\omega') & = & \sqrt{\frac{2N}{\pi\delta \Delta }}\exp \left(
\frac{(\o+\o'-\o_p)^2}{2\delta^2}-\frac{(\o-\o')^2}{2\Delta^2} \right).
\end{eqnarray}
These expressions can be justified on the grounds of the first-order perturbation theory \cite{WasilewskiXXX05}, and
for simplicity we have assumed that the phase factors resulting from linear dispersive propagation have been
compensated. The formula for $S(\omega,\omega')$ involves $\omega_p$ as the central frequency of the pump pulse,
$\delta$ and $\Delta$ describing the frequency correlations within the photon pair as well as the biphoton bandwidth,
and $N \ll 1$ equal to the average total number of generated photons. In a typical arrangement $\delta \ll \Delta$ and
$\delta$, given by the bandwidth of the pump pulse, defines the frequency anticorrelations within a biphoton, while
$\Delta$ is determined by the crystal dispersion across the down-converted spectrum and specifies the overall bandwidth
of the parametric fluorescence.

The Gaussian form allows us to find analytically \cite{BanaszekAPS} the singular values as
\begin{equation}
\label{Eq:zetanGaussian}
 \zeta_n \approx \sinh \zeta_n = \frac{\sqrt{N}}{\cosh r} \tanh^n r
\end{equation}
where $r=\ln\sqrt{\Delta/\delta}$, while the corresponding characteristic functions are given by:
\begin{equation}
\label{Eq:PsiHermite} \phi_n(\o) = \psi_n(\o) = \sqrt{\frac{\tau_s}{2^n n! \sqrt{\pi}}} H_n(\tau_s (\omega -
\omega_p/2)) \exp\left[- \frac{\tau_s^2}{2} \left(\o - \frac{\o_p}{2}\right)^2 \right]
\end{equation}
where $H_n(x)$ denotes the $n$th Hermite polynomial. Thus in our model the characteristic modes have the familiar shape
of Hermite functions describing the eigenfunctions of a harmonic oscillator, with their width given by the parameter $\tau_s$:
\begin{equation}
\tau_s = \sqrt{\frac{2}{\delta\Delta}}.
\end{equation}
Let us now assume that the local oscillator has a Gaussian spectral profile characterized by a bandwidth $\delta_{LO}$:
\begin{equation}
\label{Eq:psiLO}
\psi_{LO}(\omega) =  \sqrt{\frac{2}{\delta_{LO}\sqrt{\pi}}}
\exp \left[-\frac{1}{\delta_{LO}^2}
\left(\omega - \frac{\omega_p}{2} \right)^2 \right].
\end{equation}
The local oscillator mode can be decomposed in the orthonormal basis of the characteristic output modes as:
\begin{equation}
\label{Eq:psiLOdecomp}
\psi_{LO}(\omega) = \sum_{n=0}^{\infty} M_{n} \psi_{n}(\omega).
\end{equation}
With the choice of the spectral phase as in Eq.~(\ref{Eq:psiLO}) the local oscillator pulse is matched temporally with the
squeezed light, and only even characteristic squeezing modes appear in the decomposition owing to the defined parity of the
modes. Furthermore, the expansion coefficients in Eq.~(\ref{Eq:psiLOdecomp}) can be evaluated analytically \cite{Yuen76}:
\begin{equation}
M_{2m}=
\int\text{d}\omega \, \psi_{LO}(\omega) \psi^{\ast}_{2m}(\omega)
=\frac{\sqrt{(2m)!}}{2^{m} m!} \frac{\tanh^{m} r'}{\sqrt{\cosh r'}}
\end{equation}
where $r'=\ln(\delta_{LO}/\sqrt{\delta\Delta})$, and all $M_{2m+1}=0$.
As the modes $\psi_n(\omega)$ are mutually orthogonal, the quadrature fluctuations
detected in the local oscillator mode $\psi_{LO}(\omega)$ are given by a sum of independent contributions from all the
occupied characteristic modes.
In each of these modes, the mean square quadrature noise is phase dependent, with the
maximum $\langle Q_{n+}^2\rangle$ and the minimum $\langle Q_{n-}^2\rangle$ given by
\begin{equation}\langle Q_{n\pm }^2\rangle = \frac{1}{4} \exp(\pm
2\zeta_n),\label{Qm2}
\end{equation}
where the vacuum state noise is normalized to $1/4$. Because a pure squeezed state is a minimum uncertainty state, the
product of the minimum and maximum quadrature uncertainties is equal to that of the vacuum. In the model Gaussian case only even
characteristic modes contribute, and furthermore the coefficients $M_{2m}$ are real and thus have identical phases.
Consequently we can set the phase of the local oscillator to a value which
combines contributions from exclusively either
squeezed or antisqueezed quadratures. Thus the fluctuations in the squeezed
$\langle Q_{-}^2 \rangle$ and the antisqueezed $\langle Q_{+}^2 \rangle$ quadratures are given by:
\begin{equation}
\label{QLO2}
\langle Q_{\pm}^2 \rangle = \frac{1}{4} \sum_{m=0}^{\infty} M_{2m}^2 e^{\pm 2\zeta_{2m}}.
\end{equation}
We see that the squeezing is still present, but is not necessarily of a minimum-uncertainty character. Uncertainty
minimization occurs only if all the $\zeta_n$, for which $M_n$ do not vanish, are equal. The latter requirement is
fulfilled when the local oscillator mode is identical to one of the characteristic modes. However, even if this is not
the case, minimum uncertainty is approximated remarkably well as long as the dominant terms in the series $\{M_{2m}\}$
pick up characteristic modes with comparable squeezing parameters $\zeta_{2m}$.

\begin{figure}[!t]
    \begin{center}
        \includegraphics[width=0.45\textwidth]{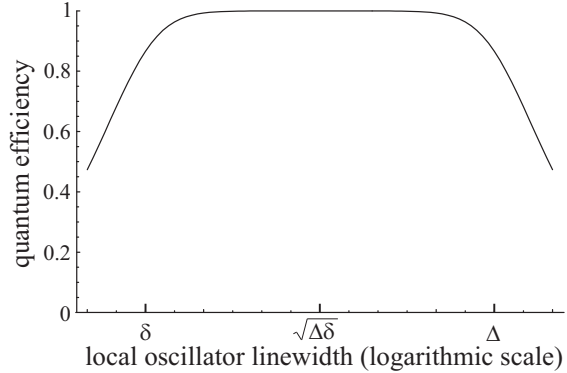}
        \caption{Quantum efficiency of detected squeezing as
        a function of the local oscillator linewidth $\delta_{LO}$ for
        $r=\ln \sqrt{\Delta/\delta}=3$. High quantum efficiency is
        observed for all $\delta_{LO}$s between $\delta$ and
        $\Delta$, with $\eta>99$ \% for $-2<r'<2$.}
        \label{Fig:eta}
    \end{center}
\end{figure}

In order to quantify the last statement, we make use of the fact that a deviation of a squeezed ensemble from the
minimum-uncertainty character can be ascribed to a non-unit quantum efficiency
\begin{equation} \eta = \frac
{-16\langle Q_+^2\rangle\langle Q_-^2\rangle+4\langle Q_+^2\rangle + 4\langle Q_-^2\rangle-1}{4\langle
Q_+^2\rangle+4\langle Q_-^2\rangle-2}. \label{eta} \end{equation}
of the communication channel along which it has been
delivered. Interpreting Eq.~(\ref{QLO2}) in this fashion, we find the efficiency of the squeezed ensemble in the local
oscillator mode. This quantity is plotted in Fig.~\ref{Fig:eta}, where it is seen that high-efficiency squeezing is
obtained in a broad span of modes. The spectral width of squeezed modes is limited from below by the master laser
linewidth and from above by the phase matching range of the down-conversion arrangement. By choosing a sufficiently
short crystal, one can control the upper limit, allowing generation of squeezed vacuum in arbitrarily short pulsed
modes. Our results permit an intuitive interpretation if down-conversion is considered in the time representation. The
two photons in a pair can be born at any moment within the duration of the pump pulse, but necessarily at the same
time. Therefore, all the infinitely short pulsed modes within the pump pulse must contain an even number of photons and
are squeezed, and so are all the linear combinations of these modes --- that is, all the modes that are shorter than
the pump pulse. The existence of the upper limit $\Delta$ in the squeezing spectrum is explained by the nonzero length
of the down-conversion crystal. Due to nonlinear dispersion (which limits the down-conversion spectrum), single-photon
wavepackets propagating through the crystal diverge in time by $\Delta^{-1}$. If the local oscillator pulse is chosen
too short, it may happen that one photon in a pair is registered within the local oscillator mode, but the other one
arrives either too soon or too late, leading to a reduction in detection efficiency.

\begin{figure}[!t]
    \begin{center}
        \includegraphics[width=0.4\textwidth]{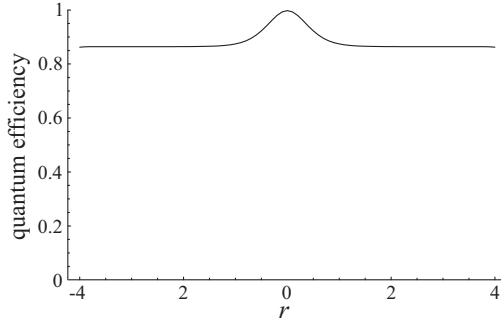}
        \caption{Quantum efficiency of a pulsed squeezed state for
        the local oscillator pulse bandwidth equal to that of
        the master laser, depicted as a function of the parameter $r$
        characterizing frequency correlations.}
        \label{Fig:1MasterLaser}
    \end{center}
\end{figure}

Most existing experiments on pulsed squeezing obtain the local
oscillator from the master laser whose second harmonic serves as a
pump for the down-conversion. Assuming that the second harmonic
bandwidth is limited by the spectrum of the pump pulse,
this situation corresponds to
$\delta_{LO} = \delta$ and consequently $r'=-r$. From Fig.~\ref{Fig:eta} we find that such a
choice of the local oscillator linewidth lies on the border of the
high-efficiency region and is not optimal. Fig.~\ref{Fig:1MasterLaser} displays the
quantum efficiency associated with such a measurement as a
function of $r$. For small $r$'s, this quantity approaches unity,
but for highly eccentric biphoton spectra it appears to tend to a
constant value of $0.86$.

\section{Numerical results}

In this section we will discuss the non-perturbative regime for a realistic example of the parametric process in a
nonlinear waveguide. For this purpose we will solve numerically one-dimensional propagation, described by the set of
differential equations for the annihilation operators of monochromatic modes $\hat{a}(\o;z)$ parameterized with the
propagation distance $z$ in the medium:
\begin{equation}
\label{Eq:dadz}
\frac{\partial \hat a(\omega;z)}{\partial z} = i k(\omega) \hat a(\omega;z)
 + \frac{1}{ \LNL E_0} \int \text{d}\omega'\,e^{i
k_p(\omega'+\omega)z}E_p(\omega'+\omega)\hat a^\dagger(\omega';z).
\end{equation}
The first term on the right-hand side represents the linear propagation of the field in the medium, while the second
one is responsible for the nonlinear interaction. In the above expression, $E_p(\omega)$ is the spectral amplitude of
the pump field, and $k(\omega)$ and $k_p(\omega)$ are respectively the wave vectors of the signal and the pump fields.
We have rescaled the spectral amplitude of the pump field by $E_0=\int \text{d}\omega \, E_p(\omega)$, which allows us to
introduce a single parameter $\LNL$, expressed in the units of length, that characterizes the strength of nonlinear
interaction. In particular, $\LNL$ is inversely proportional to the amplitude of the pump field. In the case of a monochromatic pump and no phase mismatch
between the pump and the signal fields, $\LNL$ is a distance over which the root mean square quadrature noise scales by a factor of
$e$.

\begin{figure}
  \center
  \begin{tabular}{cc}
    $C(\o,\o')$ & $S(\o,\o')$ \\
    \includegraphics[width=0.43\textwidth]{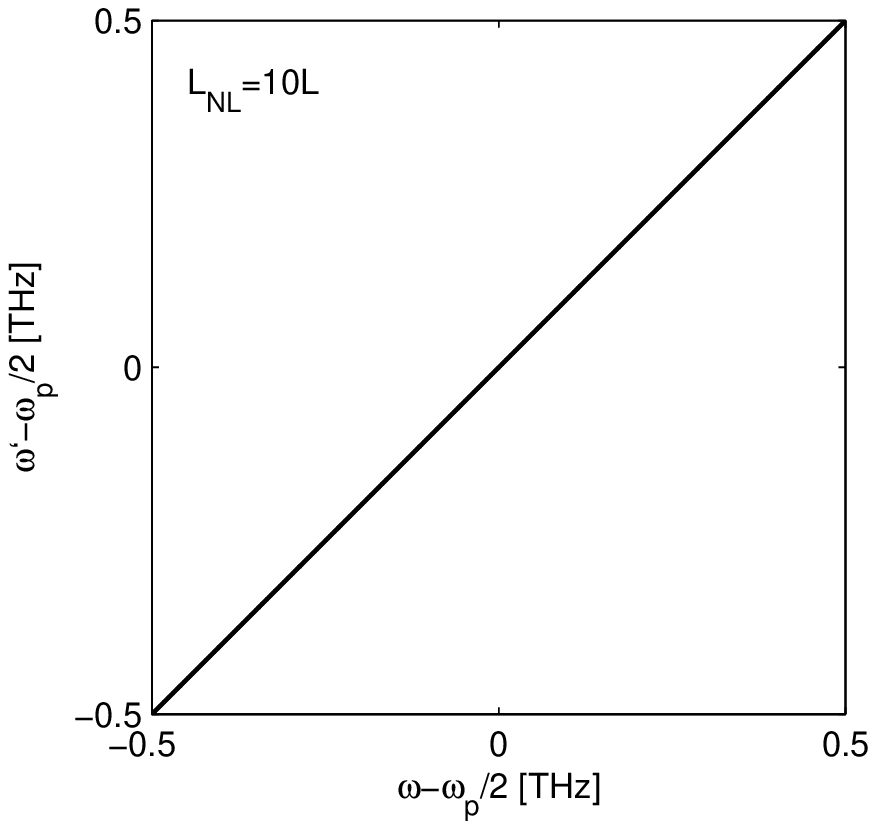} & \includegraphics[width=0.43\textwidth]{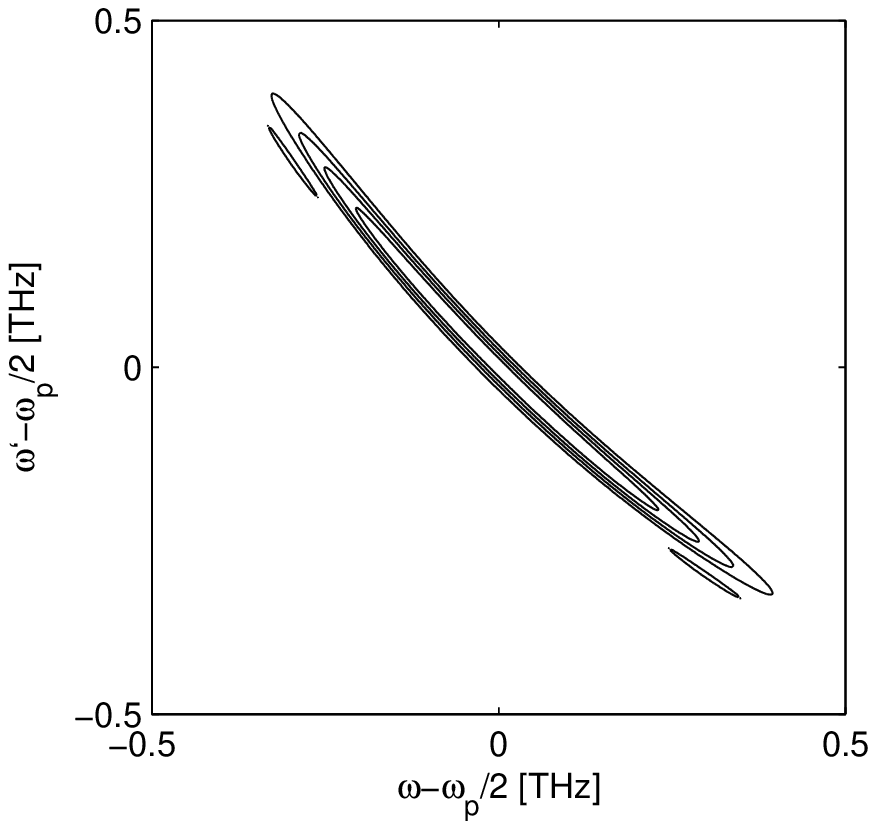}\\
    \includegraphics[width=0.43\textwidth]{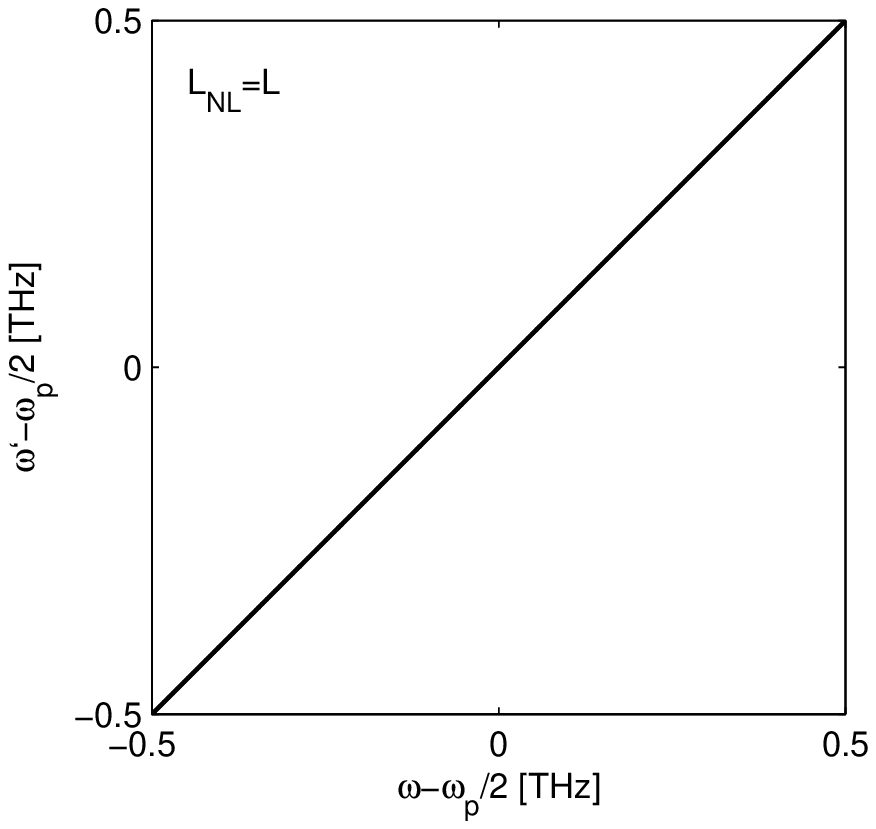} & \includegraphics[width=0.43\textwidth]{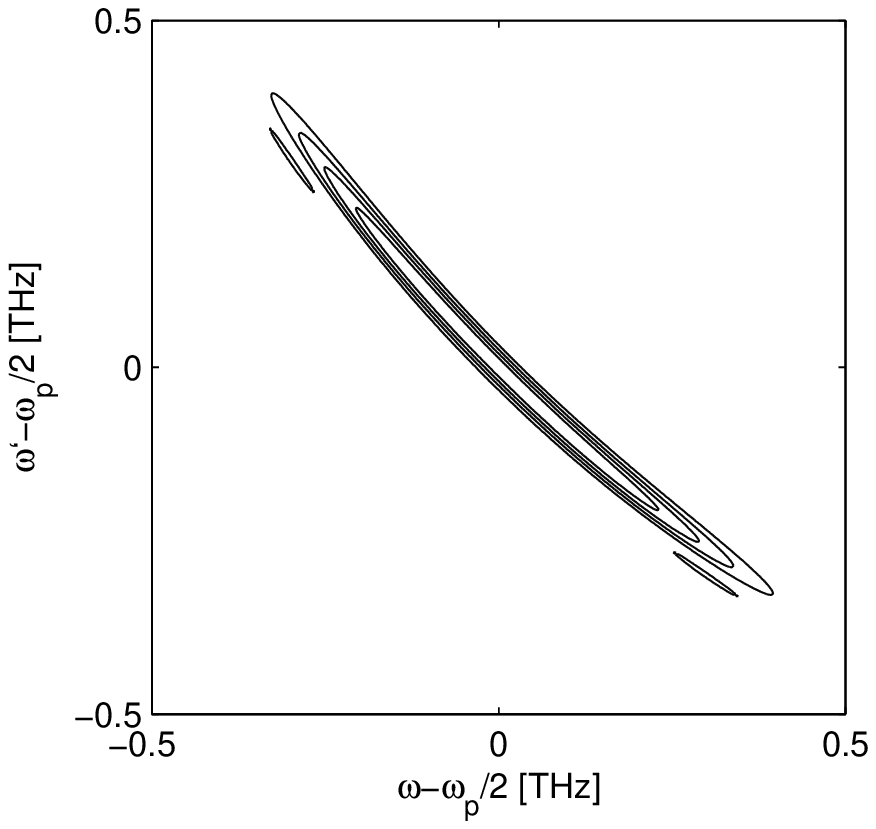}  \\
    \includegraphics[width=0.43\textwidth]{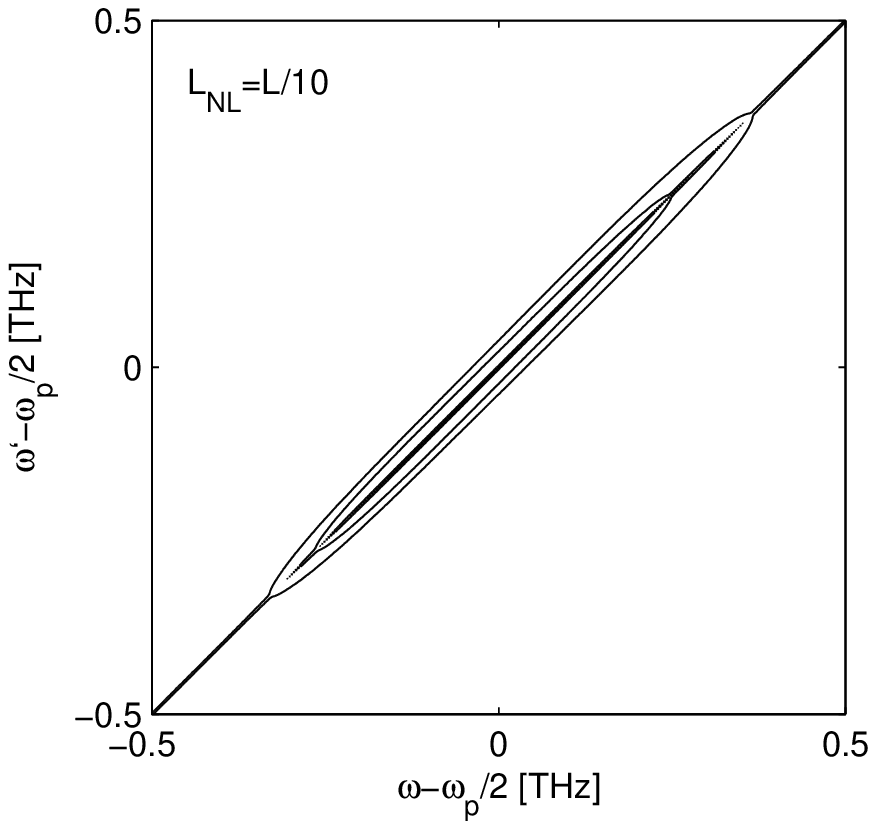} & \includegraphics[width=0.43\textwidth]{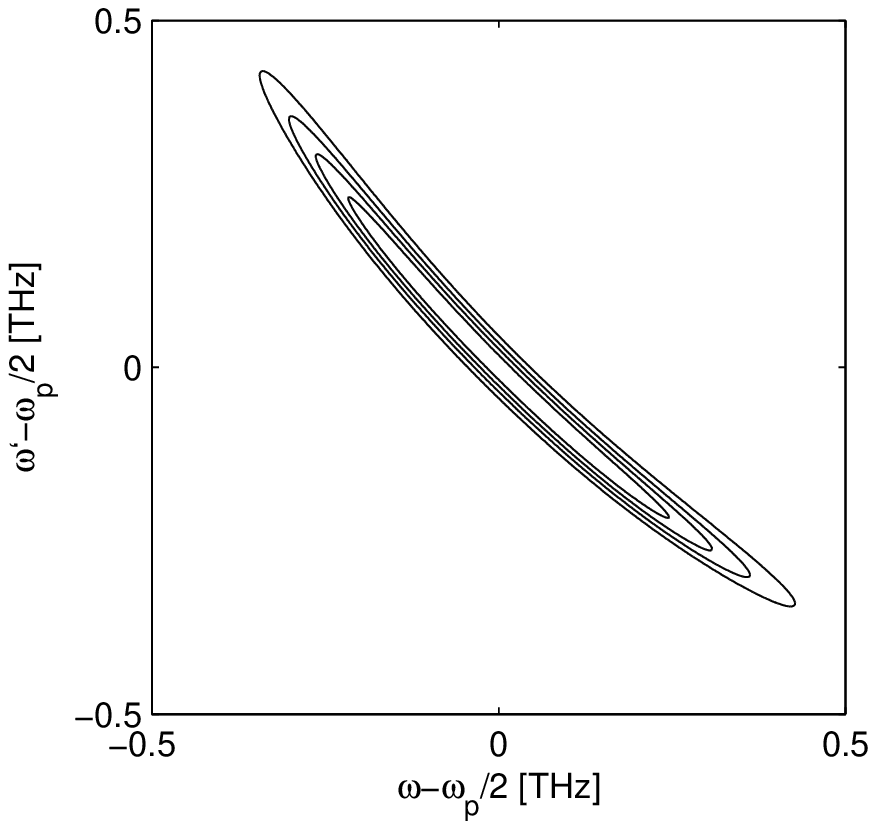}  \\
  \end{tabular}
  \caption{Green functions $|C(\o,\o')|$ and $|S(\o,\o')|$ for the example of a 1~mm long waveguide in a BBO crystal
  and the increasing ratio $L/\LNL=0.1$ (upper pair), $L/\LNL=1$ (centre pair), and $L/\LNL=10$ (lower pair). The pump
  pulse duration is equal to $\tau_p=26$~fs and its central wavelength is 400~nm. Contours are drawn at 0.2, 0.4, and 0.6 of the maximum value.}
  \label{Fig:GreenFncts}
\end{figure}

We have performed calculations for a down-conversion process taking place in a beta-barium borate (BBO) crystal and
converting pump wavelengths centered at 400~nm to degenerate signal and idler wavelengths around 800~nm. The pump pulse
was assumed to have a Gaussian shape of duration $\tau_p$, corresponding to the spectral amplitude:
\begin{equation}\label{eq:pump}
E_p(\omega) \propto \exp\left(- \frac{\tau_p^2 }{2} (\omega-\omega_p)^2 \right)
\end{equation}
with a flat spectral phase at half-way through the waveguide length. As the propagation equation (\ref{Eq:dadz}) is
linear in the field operators, it can be solved numerically by replacing $\hat{a}(\o;z)$ and $\hat{a}^\dagger(\o;z)$ by
c-numbers and correspondingly their complex conjugates. The complete Green functions are composed of solutions of
the propagation equation for monochromatic inputs. Details of the numerical procedure have been described in
Ref.~\cite{WasilewskiXXX05}. In Fig.~\ref{Fig:GreenFncts} we depict the absolute values of the Green functions
$|C(\omega,\omega')|$ and $|S(\omega,\omega')|$ for several ratios $L/\LNL$, with the length of the crystal $L=1$~mm,
and the pump pulse duration $\tau_p = 26$~fs. It is seen that the analytical Gaussian form of the Green functions
discussed in the preceding section should be indeed capable of modelling qualitatively the numerical results.

\begin{figure}
  \center
  \includegraphics[width=0.32\textwidth]{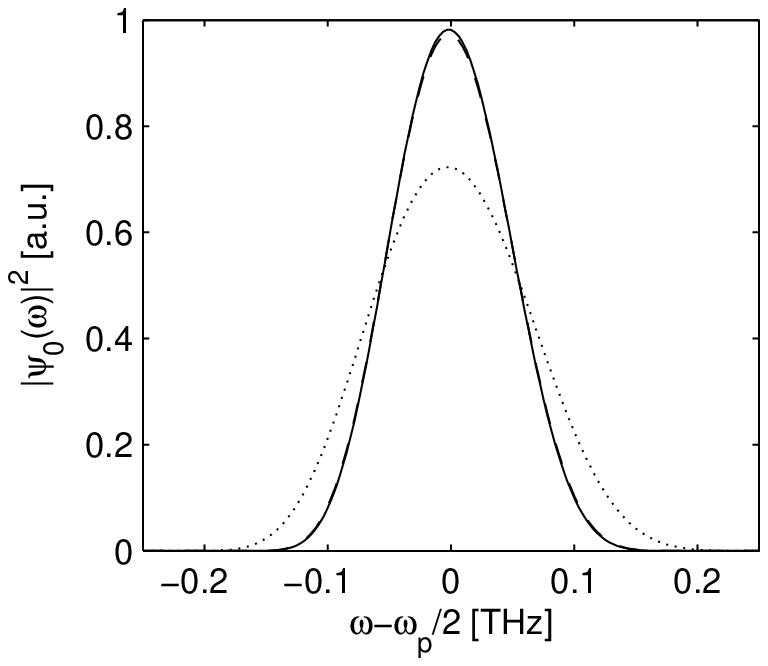}
  \includegraphics[width=0.32\textwidth]{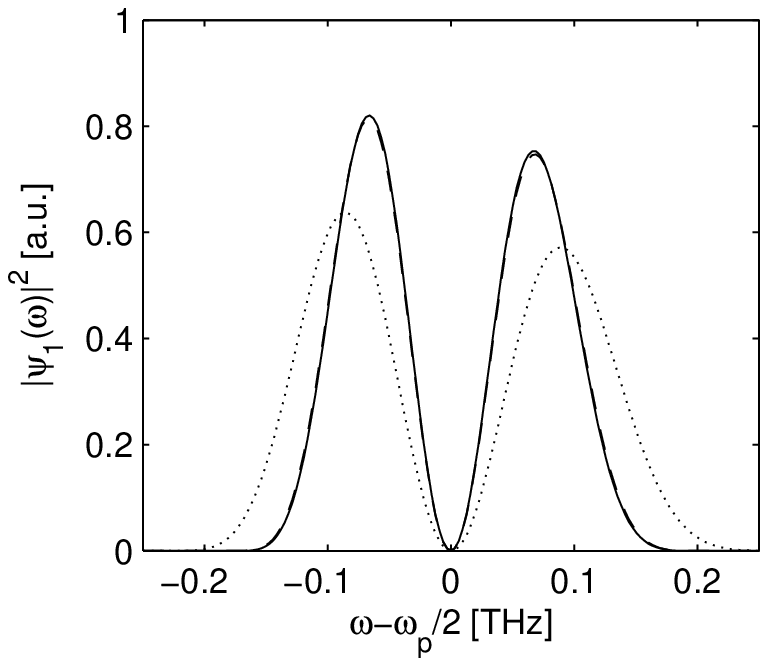}
  \includegraphics[width=0.32\textwidth]{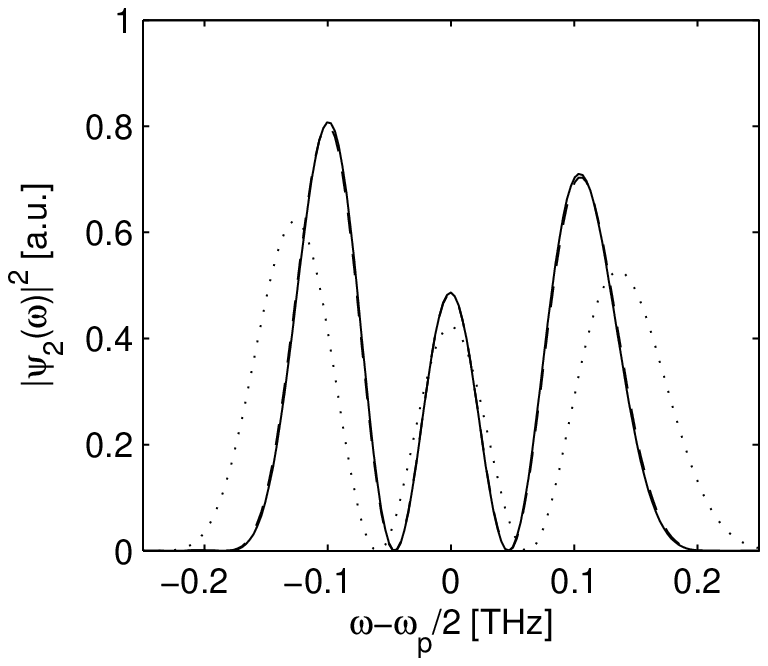}
  \caption{Spectral intensity profiles of the characteristic modes exhibiting strongest squeezing for
  interaction the lengths $L/\LNL=0.1$ (solid), $L/\LNL=1$ (dashed), and $L/\LNL=10$ (dotted). Note that the dashed and the solid lines virtually overlap in the graphs.}
  \label{Fig:chmodes}
\end{figure}

Let us begin by inspecting the spectral intensity profiles for the characteristic modes $\psi_n(\omega)$. The examples
shown in Fig.~\ref{Fig:chmodes} demonstrate that the dependence of the profiles on the nonlinear interaction length is
rather weak. Furthermore, it is easy to recognize the similarity of the intensity profiles to those described by the
Hermite functions given in Eq.~(\ref{Eq:PsiHermite}). Nevertheless small deviations from the Hermitian shapes,
especially breaking the defined parity of the characteristic modes, can have a dramatic impact on the amount of
squeezing detectable with a restricted class of local oscillator shapes. The reason for this is that when detecting the
squeezed quadrature, such local oscillator pulses would pick up an admixture of antisqueezed quadratures from odd
characteristic modes, which can mask the effect of squeezing. It is also difficult to set all the phases in the
characteristic-mode decomposition to the same value which selects only squeezed quadratures. This issue has been
discussed in detail in Ref.~\cite{WasilewskiXXX05}.

\begin{figure}
  \center\includegraphics[width=0.45\textwidth]{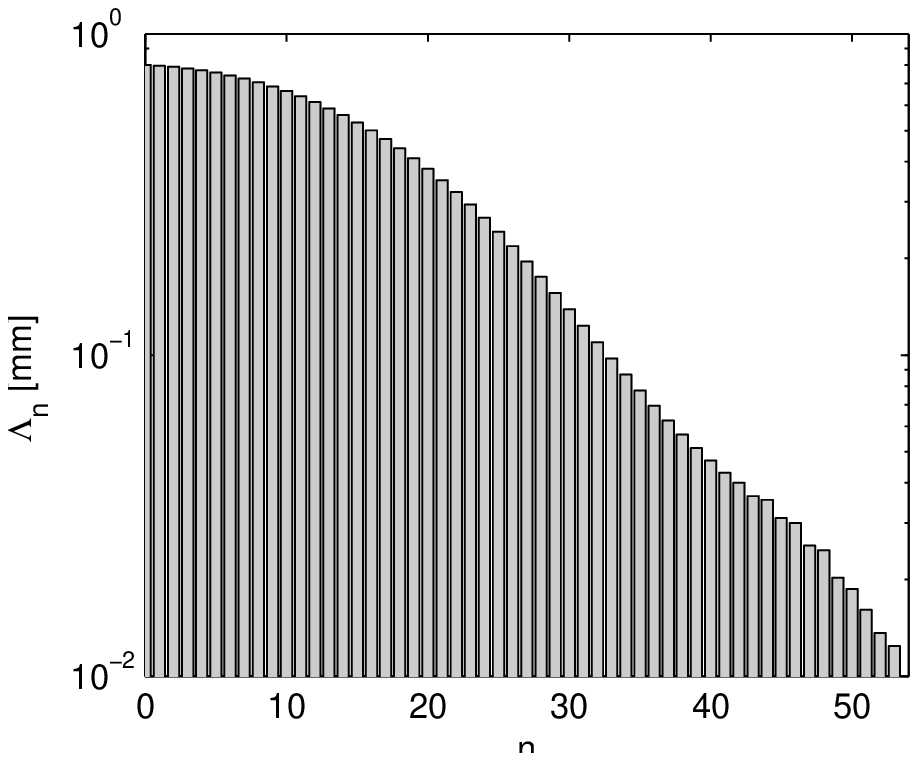}
  \caption{The squeezing lengths $\Lambda_{n}=\zeta_n\LNL$ as a function of the mode number $n$ for the validity range of the inverse linear scaling law, $0 \le L/\LNL \le 15$.}
  \label{gaind}
\end{figure}

The squeezing experienced by the $n$th characteristic mode is given by the squeezing parameter $\zeta_n$.
 We found numerically that up to $L/\LNL=15$
there exist a simple scaling law: namely, that the squeezing parameter of any squeezed mode is inversely
proportional to nonlinear length
\begin{equation}\label{eq:lambdasq}
\zeta_n=\frac{\Lambda_{n}}{\LNL}.
\end{equation}
We have introduced here a proportionality constant $\Lambda_{n}$ which we shall call the squeezing length. In order to
discuss its physical meaning, let us first recall that in the case of a monochromatic pump and no phase mismatch
between the pump and the signal fields, we would expect the squeezing parameter to be equal to the ratio of the interaction length to
the nonlinear length $\zeta=L/\LNL$. However in the general multimode case studied here there is an appreciable phase
mismatch and the pump field is a polychromatic ultrashort pulse. Consequently the squeezing parameters cannot be simply
estimated as $L/\LNL$. Nevertheless, Eq.~(\ref{eq:lambdasq}) suggests that one can introduce a notion of the {\em
effective crystal length} $\Lambda_{n}$ for every characteristic mode, a quantity that depends on the mode number as
well as on the pump and the phase matching properties of the medium. We propose the following way of intuitively
understanding $\Lambda_{n}$: for a given input mode shape $\phi_n(\o)$ its interaction with the pump
pulse remains phase matched, thus interacting efficiently, only over a finite length of the
crystal, approximately equal to $\Lambda_{n}$. This length is shorter for modes of broad spectral content as their
various frequency components dephase more quickly from each other and from the pump pulse owing to crystal dispersion.
These effects are clearly seen in Fig.~\ref{gaind}, where we plot $\Lambda_{n}$ as a function of the mode number $n$.
Indeed, $\Lambda_{n}$s are smaller for modes with higher $n$, whose bandwidth is broader. Let us point out here that
the Gaussian approximation fails to predict accurately the squeezing parameters at any $\LNL$, as the plot of the analytical values given by
\eeqref{Eq:zetanGaussian} would be a straight line
in the semi-logarithmic scale of Fig.~\ref{gaind}.
This is a consequence of the fact that even in the perturbative regime the biphoton spectral amplitude is not Gaussian, but typically involves a much more slowly decaying sinc function.
Finally, let us note that first several modes have very similar squeezing lengths $\Lambda_{n}$ and thus squeezing
parameters $\zeta_n$. As discussed in the preceding section, this gives us freedom in the shapes of the local
oscillator pulse that are capable of detecting maximum squeezing.

\section{Conclusions}

The Bloch-Messiah reduction is an indispensable theoretical tool in understanding the intricacies of optical parametric
amplification in the broadband regime, when all the monochromatic modes are coupled to one another through nonlinear
interaction with the ultrashort pump pulse. The characteristic modes are a simple and elegant way to describe
completely the quantum statistical properties of parametric fluorescence. They also form a convenient modal
decomposition to analyse homodyne detection of squeezing.

Numerical simulations presented in this paper show that the perturbative approach begins to fail when the ratio
$L/\LNL$ approaches one. This threshold value corresponds to quadrature squeezing exceeding the order of 1/e, which is
equivalent to over 8~dB in noise reduction. This suggests that the perturbative approximation should be quite
sufficient for describing squeezing experiments performed so far. As long as we stay within the perturbative regime,
the Gaussian model developed in Sec.~II shows that there is a certain degree of flexibility in selecting the bandwidth
of the local oscillator, with the optimal range limited by the parameters $\delta$ and $\Delta$ characterizing
frequency correlations within biphotons. However, when we utilize the whole available bandwidth of the master laser for
second-harmonic generation, extracting the local oscillator from the same master beam puts us with $\delta_{LO}$ close
to $\delta$, which is right on the edge of the optimal region. One possible remedy is to limit the bandwidth of the
pump pulse by generating second harmonic in a longer crystal, in the dispersion-limited regime. This would lower the
value of $\delta$ and place the master laser badwidth $\delta_{LO}$ used for homodyning inside the favourable region.

\section*{Acknowledgements}
We acknowledge helpful discussions with C. Radzewicz, M. G. Raymer, I. A. Walmsley, and K.-P. Marzlin, as well as
financial support from CFI, NSERC, CIAR, AIF, MNiI grant number 2P03B 029 26, and the European Commission through Integrated Project QAP (contract 015848).

\end{document}